\documentclass{elsart}
\usepackage{graphicx}
\usepackage{amsmath}
\usepackage[usenames]{color}
\usepackage{lineno}

\usepackage{amssymb}

\usepackage[english,francais]{babel}

\newtheorem{e-proposition}[theorem]{Proposition}

\newtheorem{e-definition}[theorem]{Definition\rm}

\renewcommand{\theequation}{\arabic{equation}}
\def\og{\leavevmode\raise.3ex\hbox{$\scriptscriptstyle\langle\!\langle$~}}
\def\fg{\leavevmode\raise.3ex\hbox{~$\!\scriptscriptstyle\,\rangle\!\rangle$}}

\def\deg{{^\circ}}
\newcommand{\Ro}{\text{Ro}}
\newcommand{\E}{\text{E}}
\newcommand{\Rm}{\text{Rm}}

\newcommand{\Ha}{\text{Ha}}
\renewcommand{\Re}{\text{Re}}
\newcommand{\N}{\text{N}}

\begin{document}
% Select a primary header Physics or Astrophysics
% You can place after the header (classification), if you know it.

\centerline{Physics or Astrophysics/Header}
\begin{frontmatter}


% \author{Name\thanksref{label2}}
\selectlanguage{english}
\title{On the peculiar nature of turbulence in planetary dynamos}

% \author[label1,label2]{Henri-Claude Nataf}
% \author{Nad\`ege Gagni\`ere}
% If all authors are at the same address, the [label1] can be suppressed

\selectlanguage{english}
\author{Henri-Claude Nataf},
\ead{Henri-Claude.Nataf@ujf-grenoble.fr}
\author{Nad\`ege Gagni\`ere}
\ead{Nadege.Gagniere@ujf-grenoble.fr}

\address{University of Grenoble - Centre National de la Recherche Scientifique, LGIT, BP 53, Maison des Geosciences, 38041 Grenoble cedex 9, France}
%\address[authorlabel2]{Address2}

% If you know the dates of reception, and acceptation you can put them now;
%    idem the name of the person presenting your article

\medskip
\begin{center}
{\small Received *****; accepted after revision +++++}
\end{center}

\begin{abstract}
Under the combined constraints of rapid rotation, sphericity, and magnetic
field, motions in planetary cores get organized in a peculiar way. Classical
hydrodynamic turbulence is not present, but turbulent motions can take place
under the action of the buoyancy and Lorentz forces. Laboratory experiments,
such as the rotating spherical magnetic Couette {\it DTS} experiment in Grenoble,
help us understand what motions take place in planetary core conditions.
{\it To cite this article: H.-C. Nataf and N. Gagni\`ere, C. R.
Physique XXX (2008).}

\vskip 0.5\baselineskip

\selectlanguage{francais}
\noindent{\bf R\'esum\'e}
\vskip 0.5\baselineskip
\noindent
{\bf Sur la nature particuli\`ere de la turbulence dans les noyaux plan\'etaires. }
Sous les contraintes combin\'ees de la rotation rapide, de la sph\'ericit\'e et du champ magn\'etique, les
\'ecoulements dans les noyaux plan\'etaires s'organisent d'une mani\`ere particuli\`ere. La turbulence
hydrodynamique classique n'est pas pr\'esente mais des mouvements turbulents peuvent se mettre en place
sous l'action des forces d'Archim\`ede et de Lorentz. Des exp\'eriences de laboratoire comme l'exp\'erience
{\it DTS} de Couette sph\'erique sous champ magn\'etique \`a Grenoble, nous aide \`a comprendre les \'ecoulements
qui peuvent exister dans les conditions des noyaux plan\'etaires.
{\it Pour citer cet article~: H.-C. Nataf and N. Gagni\`ere, C. R.
Physique XXX (2008).}

%Now keywords/mots-clÈs
 \vskip 0.5\baselineskip
\noindent{\small{\it Mot-cl\'es~:} Dynamo~; Noyau plan\'etaire~; DTS; Couette sph\'erique}}
\end{abstract}
\end{frontmatter}

% now the Version franÁaise abrÈgÈe, if it exists
%\selectlanguage{francais}
%\section*{Version fran\c{c}aise abr\'eg\'ee}
% Text of your Version franÁaise abrÈgÈe here
% La turbulence dans le noyau de la Terre et des plan\`etes... 

\selectlanguage{english}
% main text
\section{Introduction}
\label{introduction}

The recent success of the {\it VKS} dynamo experiment \cite{Monchaux06,Berhanu07}, after the harvest of the
pioneer experiments in Riga \cite{Gailitis01} and Karlsruhe \cite{Stieglitz01}, brings hope to better understand the mechanisms at work
in the dynamo process. It is believed that the self--sustained dynamo process generates the magnetic fields
of most astrophysical objects, including our planet, the Earth. In this process, a given velocity field $\vec{u}$
can produce a magnetic field controlled by the magnetic induction equation (\ref{eq:induction}), which shows that
if the velocity is large enough for the induction term $\vec{\nabla} \times \left( \vec{u} \times \vec{B} \right)$ to
dominate over the diffusion term $\eta \Delta \vec{B}$, the magnetic field $\vec{B}$ can grow.
%---------------
\begin{equation}
\frac{\partial \vec{B}}{\partial t} = \vec{\nabla} \times \left( \vec{u} \times \vec{B} \right) + \eta \Delta \vec{B}
\label{eq:induction}
\end{equation}
%---------------
The magnetic field of the Earth originates in its core, made of liquid iron. As the Earth cools down, on a geological time--scale, heat is carried
from the core to the mantle by convective motions. These motions are governed by the Navier--Stokes equation, which, in the Boussinesq--approximation
may be written as:
%---------------
\begin{equation}
\rho \left( \frac{\partial \vec{u}}{\partial t} + \vec{u} \cdot \nabla \vec{u} \right) = - \vec{\nabla} P - 2 \rho \vec{\Omega} \times \vec{u} + \mu \Delta \vec{u} + \vec{j} \times \vec{B} + \vec{f}_a
\label{eq:NS}
\end{equation}
%---------------
where the various symbols have their usual meaning. On the right hand side, one recognizes the Coriolis force $-2 \vec{\Omega} \times \vec{u}$, the
viscous force $\mu \Delta \vec{u}$, the Lorentz force $\vec{j} \times \vec{B}$, and the buoyancy force $\vec{f}_a$.

When plugging in typical values of the properties of liquid iron and of the velocity $\vec{u}$ and magnetic $\vec{B}$ 
fields, one finds that, because of the fast rotation $\Omega$, the dominant forces are the Coriolis and Lorentz
forces. The resulting regime is called ``magnetostrophic'', and it has received the attention of theoreticians
for decades. Taylor \cite{Taylor63} was the first to point out that, in this regime, if one considers tubes of liquid
co--axial with the rotation axis, there is no torque to resist the torque applied by the Lorentz forces. He concluded that,
in the steady--state, the torque applied by the Lorentz forces must vanish, yielding what is now called a ``Taylor--state''.
The magnetostrophic regime and the expected Taylor--state remain very difficult to reach in numerical simulations (because it is technically difficult
to neglect viscous terms) and their exploration in laboratory experiments is recent \cite{Nataf06,Schmitt08,Nataf08}.

In this article, we will highlight some striking properties of the ``magnetostrophic'' regime, as derived
from measurements performed in the {\it DTS} experiment. In section \ref{DTS}, we describe the {\it DTS} setup and
present the 
relevant parameters. The time--averaged flow is analyzed in section \ref{mean_flow}, and the (weak) fluctuations in
section \ref{weak_turbulence}. Implications for turbulence in planetary cores is discussed in section \ref{core_turbulence}.

%------------------------------------------------------------------------
\section{The {\it DTS} experiment: exploring the magnetostrophic regime}
\label{DTS}
%------------------------------------------------------------------------

\begin{figure}
  \centerline{ \includegraphics[width=10cm]{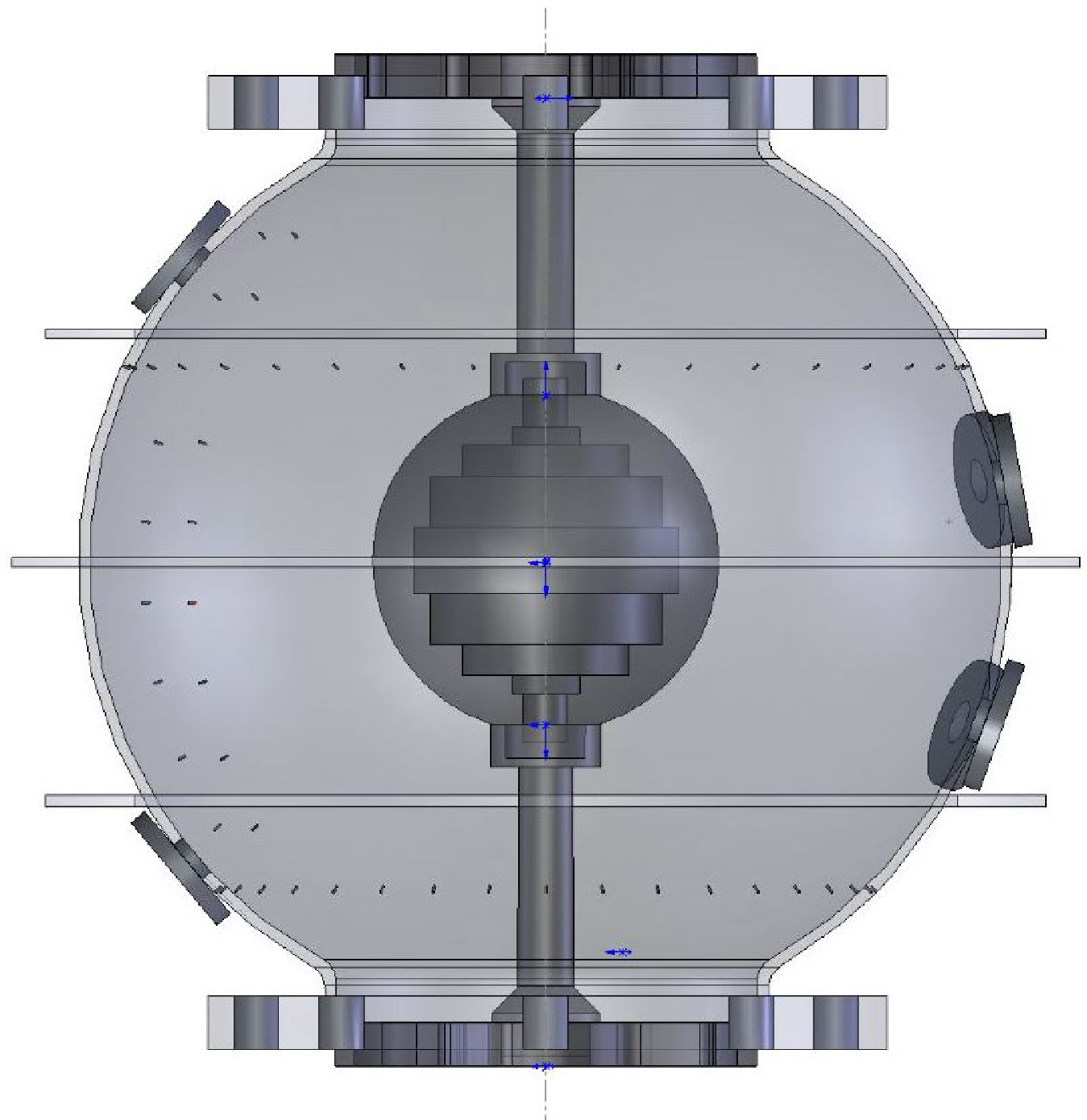}}
     \caption{Sketch of the central part of the $DTS$ experiment. The $7.4 \text{cm}$--radius inner sphere is made of copper and contains 5 layers of magnets.  The inner radius of the stainless steel outer sphere is $21 \text{cm}$ and its thickness is $5 \text{mm}$. Blind holes are drilled and threaded along several meridians and parallels to receive electrodes. The large holes at latitudes $-20 \deg$, $10 \deg$ and $\pm 40 \deg$ receive interchangeable assemblies, which can be equipped with ultrasonic transducers.}
        \label{fig:sphere}
\end{figure}

The {\it Derviche Tourneur Sodium (DTS)} experiment has been designed for the exploration of the magnetostrophic regime, in which the Coriolis and
Lorentz forces are dominant \cite{Cardin02}. It consists in a spherical Couette flow where both the inner and outer spheres can rotate rapidly at
separate angular velocities around a common vertical axis. Forty liters of liquid sodium fill the shell between the two spheres and the inner sphere
contains a strong magnet. 
The set--up is depicted in figure \ref{fig:sphere}. 
The fluid flow is thus strongly influenced by both the Coriolis force and the Lorentz force. Dimensions and typical dimensionless
numbers are given in Table \ref{tab:dimen}. 

% ---------------- FIGURE SPHERE -----------------------------------------------------------
% -------------------------------------------------------------------------------------------

%--------------------------------------------------------
\begin{table}
\begin{center}
\begin{tabular}{ccccc}
\hline
symbol & expression & units  & \multicolumn{2}{c}{value} \\
\hline
 $a$  & outer radius & cm& \multicolumn{2}{c}{$21$} \\
 $b$ & inner radius & cm & \multicolumn{2}{c}{$7.4$} \\
 $\tau_J$ & $a^2/\pi^2\eta$ & s & \multicolumn{2}{c}{$0.05$} \\
 & & & \multicolumn{1}{c}{$\underline{\qquad B=B_i \qquad}$} & \multicolumn{1}{c}{$\underline{\qquad B=B_o \qquad}$}\\
 $\Ha$ & $aB/\sqrt{\mu_0\rho\nu\eta}$& &\multicolumn{1}{c}{4400} & \multicolumn{1}{c}{210}\\
 $\text{S}$ & $aB/\eta\sqrt{\mu_0\rho}$& &\multicolumn{1}{c}{12} & \multicolumn{1}{c}{0.56}\\
 $U_a$ & $B/\sqrt{\mu_0\rho}$&m s$^{-1}$ & \multicolumn{1}{c}{5.1} & \multicolumn{1}{c}{0.2}\\
   \hline
 & & & \multicolumn{2}{c}{$\underline{\qquad f = 5 \text{Hz}\qquad}$} \\
 $\E$ & $\nu/\Omega a^2$ & & \multicolumn{2}{c}{$4.7\;10^{-7}$} \\
  & & & $\underline{\: B=B_i \:}$ & $\underline{\: B=B_o \:}$ \\
 $\Lambda$ & $\sigma B^2/\rho \Omega$ & & 9.2 & 0.02 \\
 $\lambda$ & $U_a/a \Omega$ & & 0.77 & 0.04 \\
 \hline
 & & & \multicolumn{2}{c}{$\underline{\qquad \Delta f = 5 \text{Hz} \qquad}$}  \\
 $U$ & $b\Delta \Omega$ &m s$^{-1}$&\multicolumn{2}{c}{2.3} \\
 $\Rm$ & $U a/ \eta$ &&\multicolumn{2}{c}{5.5} \\
 $\Re$ & $U a/ \nu$ &&\multicolumn{2}{c}{$7\;10^5$} \\
 & & & $\underline{\: B=B_i \:}$ & $\underline{\: B=B_o \:}$ \\
 $\N$ & $\sigma a B^2/\rho U$ & & 26 & 0.006 \\
 \hline

\end{tabular}

\end{center}
\vspace{0.4cm}
\caption{Typical values of the relevant parameters and dimensionless numbers for given imposed rotation frequencies $f = \Omega/2\pi$ of the outer sphere and differential rotation of the inner sphere $\Delta f= \Delta\Omega/2\pi$ with respect to the outer sphere. For the numbers that depend on the magnetic field strength, two values are given, the first one with $B=B_i=0.175$ T at the equator of the inner sphere, the second one with $B=B_o=0.008$ T at the equator of the outer sphere.}
    \label{tab:dimen}
\end{table}
% ------------------------------------------------------------------------------------------
The radius of the outer sphere is $a = 21$cm. Because of this moderate dimension, the Joule dissipation time $\tau_J$ of
magnetic fluctuations by diffusion is small (less than one tenth of a second). The Alfv\'en wave velocity $U_a$ depends on the
strength of the magnetic field, and is thus much stronger near the inner sphere than at the outer sphere. Comparing the Alfv\'en
time to the diffusion time yields the Lundquist number S. It reaches $12$ on the inner sphere, suggesting that Alfv\'en waves
can survive there, while they must be severely damped near the outer boundary, even though the Hartmann number Ha is large
everywhere in the shell. 

The rotation frequency $f$ of the outer sphere has been varied between $0$ and $15$Hz. The small
value of the Ekman number for a typical rotation frequency of $5$Hz illustrates that the Coriolis force dominates over viscous forces
in the bulk of the fluid. A very thin Ekman layer (one tenth of a millimeter) must form beneath the outer shell. The value of
the Elsasser number $\Lambda$, which compares the Lorentz force to the Coriolis force, shows that rotation effects dominate
in that region, while the opposite holds near the inner sphere. The domain of magnetostrophic equilibrium $\Lambda \simeq 1$
occupies the central part of the fluid shell. The number $\lambda$, called Lehnert number in \cite{Jault08}, compares
the frequencies of Alfv\'en modes to that of rotation (inertial) modes. It is smaller than $1$ everywhere.

Other effects depend upon the strength of the forcing, measured by $\Delta f$, the differential rotation frequency of the inner
sphere with respect to the outer sphere. It can be varied between about $-20$Hz and $20$Hz. For a moderate value of $5$Hz,
the expected fluid velocity $U$ is of the order of $2$m/s. The Reynolds number Re is thus very large, while the magnetic
Reynolds number Rm is larger than $5$, indicating that the magnetic field is modified by the flow. The large value of
the interaction parameter N near the inner sphere demonstrates that, in turn, the magnetic field deeply influences the flow.

The $DTS$ set--up is well suited for exploring the magnetostrophic regime. It is not a dynamo experiment, but the imposed magnetic field
is strong enough to give rise to Lorentz forces of the same order as the Coriolis force. Both forces have a strong influence on the
flow. The magnetic Reynolds number is large enough for induction effects to be well developed. We now examine what flow results from
this complex combination, which is typical in planetary cores.

%------------------------------------------------------
\section{Characteristics of the mean axisymmetric flow}
\label{mean_flow}
%------------------------------------------------------

We use ultrasonic Doppler velocimetry to measure the mean axisymmetric flow. A transducer is placed in one of the interchangeable
assemblies at latitude $10 \deg$, and shoots a beam with declination and inclination angles of $60 \deg$ and $66 \deg$, respectively.
After a $38 \text{cm}$--long linear trajectory, the beam hits the outer sphere at a latitude of about $-28 \deg$. Azimuthal velocities
being more than 10 times larger than radial velocities, the along--beam measured velocity directly yields an angular velocity
profile.

\begin{figure}
  \centerline{ \includegraphics[width=14cm]{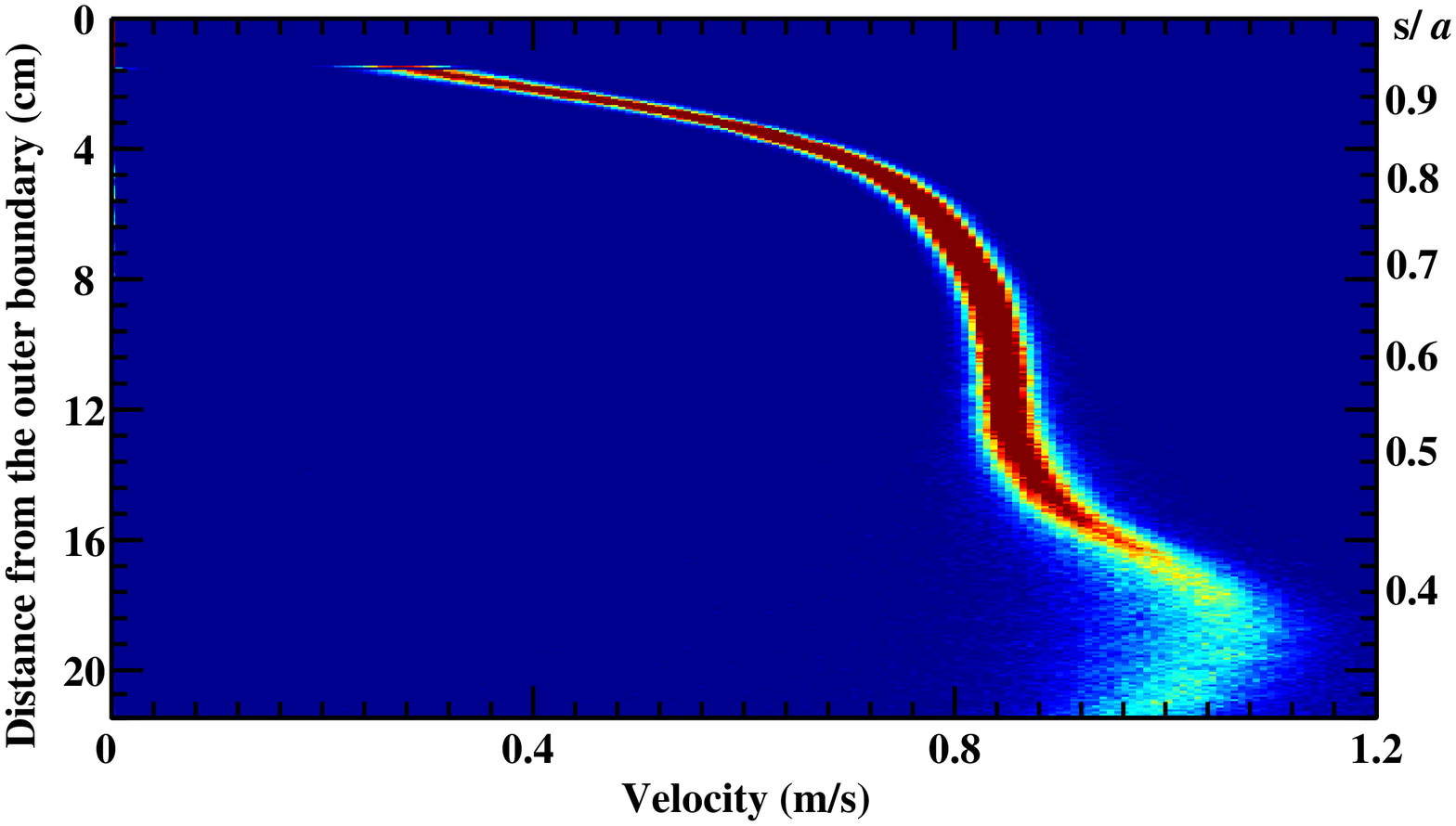}}
     \caption{Profile of the angular velocity (velocity along the beam) of the flow as a function of distance
from the outer sphere, obtained by ultrasonic Doppler velocimetry. The angular velocity rises smoothly to a plateau before
it increases again as the ultrasonic beam gets closest to the inner sphere. $f \simeq 5 \text{Hz}$ and $\Delta f \simeq 5 \text{Hz}$. The (non--linear) scale on the right gives the value of the adimensional cylindrical radius $s$. }
        \label{fig:dopazi}
\end{figure}

Such a profile is shown in figure \ref{fig:dopazi} for $f \simeq 5 \text{Hz}$ and $\Delta f \simeq 5 \text{Hz}$. Starting from the beam entry
point at latitude $10 \deg$, we see that the angular velocity rises regularly from a value of about $0.25 \text{m/s}$, reaching an almost uniform plateau
farther in the fluid, and increasing again to reach a maximum of $1.05 \text{m/s}$ where the beam gets closest to the inner sphere.
Comparing with axisymmetric numerical simulations, Nataf {\it et al.} \cite{Nataf08} explain that the velocity bump near the inner sphere
is a signature of the magnetic wind: the Elsasser number is large and the angular velocity isolines follow the magnetic field lines; the
magnetic coupling between the rotating inner sphere and the liquid sodium is efficient. Away from that bump, the flow is geostrophic
as the Elsasser number gets below $1$ (see Table \ref{tab:dimen}). The geostrophic flow is driven by the magnetic torque that results
from the shearing of the magnetic field lines and slowed down by friction in the Ekman layers beneath the outer sphere. This situation
has been well studied by Kleeorin {\it et al.} \cite{Kleeorin97}, in the asymptotic limit of small Ekman, Rossby, Reynolds and Elsasser
numbers. At small cylindrical radius $s$, the imposed magnetic field is large and friction in the Ekman layers of the outer sphere is weak. As a consequence, the shear must
be very small in order for the magnetic torque to remain small enough to balance the friction torque: the fluid is in rigid body rotation. The regular decrease in angular
velocity as one gets farther away from the inner sphere was predicted by Kleeorin {\it et al.} \cite{Kleeorin97}: friction increases
while the imposed magnetic field decreases, implying an increasing shear. It is important to realize that the flow is geostrophic,
as a consequence of strong rotation, but entirely driven by magnetic forces. A modified Taylor state is achieved, where the magnetic torque
on cylindrical tubes parallel to the axis of rotation only balances the (weak) friction in the Ekman boundary layers.
Note that, under this mechanism, the induced toroidal magnetic field is expected to be larger in the external region, where shear
is large, than in the region near the inner sphere, although the imposed magnetic field is much larger there.

We find similar profiles for all values of $\Delta f$ as long as the Rossby number $\Ro = \Delta f / f$ remains smaller than
a few units. A good quantitative fit between the theory and the measurements is achieved, when one takes into account that the Ekman
layers become turbulent in the experimental conditions (see \cite{Nataf08}). Therefore, we think that the same dynamics prevails for the very small Rossby numbers that characterize flow in the Earth's core.

% ---------------- FIGURE DOP AZIMUTHAL -----------------------------------------------------------
% -------------------------------------------------------------------------------------------

%------------------------------------------------------
\section{Weak turbulence}
\label{weak_turbulence}
%------------------------------------------------------

The profile shown in figure \ref{fig:dopazi} is, in fact, a histogram constructed from over 1000 shots.  
Although the Reynolds number reaches $7 \cdot 10^5$ for this run (see Table \ref{tab:dimen}), the fluctuations around the average profile
are amazingly small. Besides, the power spectra of the fluctuations reveal peculiar bumps at several frequencies.
This is best seen on the spectrogram of a whole run, where $f \simeq 5 \text{Hz}$ and $\Delta f$ is varied in steps from $0$ to $-20 \text{Hz}$.

% ---------------- FIGURE ddp WAVES -----------------------------------------------------------
\begin{figure}
  \centerline{ \includegraphics[width=10cm]{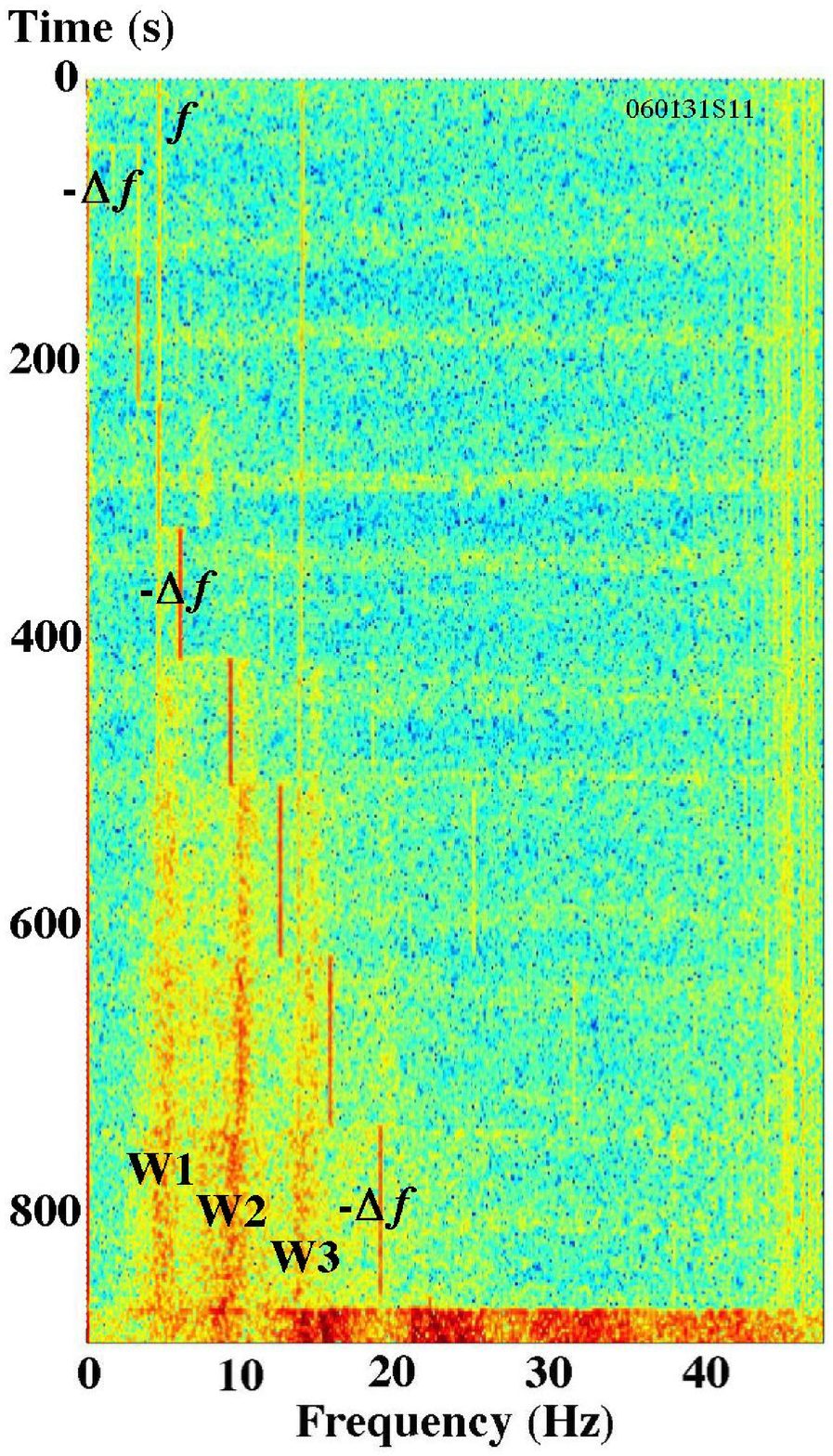}}
     \caption{Spectrogram of electric potential measurement. The signal is the electric potential measured between two
electrodes situated at latitude $-45 \deg$ and distant of $40 \deg$ in longitude. The colors give the log--amplitude of the short--time
spectral density. The horizontal axis is frequency and the vertical axis is time. During this time--frame, the outer sphere
was rotating at a constant angular frequency $f = 5 \text{Hz}$, while the inner sphere differential rotation rate $\Delta f$ was
varied in steps from $0$ to $-20 \text{Hz}$. The thin lines, labelled $f$ and $\Delta f$, reveal these frequencies and some overtones. The spectrum is almost
void until $\Delta f$ reaches about $-10 \text{Hz}$. The spectrum is then dominated by 3 broad bands (W1, W2, W3) whose frequency slightly
decreases as $|\Delta f|$ increases.}
        \label{fig:ddpwaves}
\end{figure}
% -------------------------------------------------------------------------------------------

Here, the signal is the difference in electric potential between two electrodes at the same latitude of $-45 \deg$ and distant of
$40 \deg$ in longitude (see figure \ref{fig:sphere}). The spectrogram is shown in figure \ref{fig:ddpwaves}. Frequency
is on the horizontal axis and time on the vertical axis. The power spectral density of the signal is color--coded (log--scale).
Thin lines reflect the two forcing frequencies $f$ and $-\Delta f$ and some overtones. For $-\Delta f < f$, the spectrum is featureless, while
bands of higher power are clearly visible when $\Delta f$ reaches $-10 \text{Hz}$. The detailed study of these bands performed
by Schmitt {\it et al} \cite{Schmitt08} reveal that they correspond to waves that propagate in the direction of the flow (in the
frame of reference rotating with the outer sphere) but at a slower velocity. The lowest band corresponds to a wave or mode with an azimuthal wave--number $m = 1$,
the following one to $m = 2$, and so on. The origin of these waves is not clear yet. They are different from the inertial modes identified
by Kelley {\it et al} \cite{Kelley07} in a similar set--up but with a weak magnetic field. Indeed, the derivation
of their dispersion relationship by Schmitt {\it et al} \cite{Schmitt08} reveal that their frequencies can be higher
than that of inertial modes (which are bounded by $2 f_{fluid}$, where $f_{fluid}$ is the rotation frequency of the fluid in 
the laboratory frame of reference).

\begin{figure}
  \centerline{ \includegraphics[width=10cm]{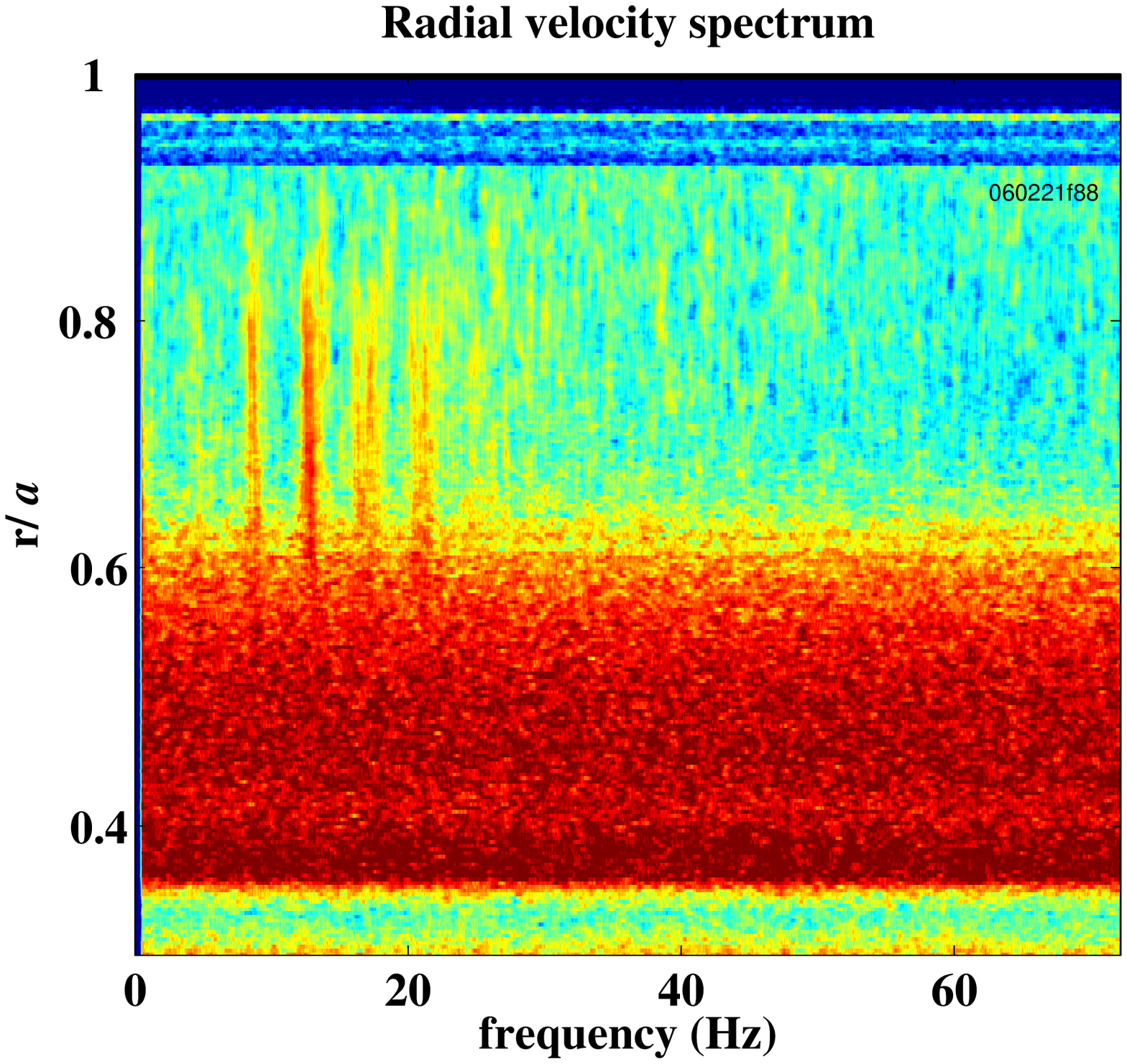}}
     \caption{Radial profile of power spectral density of radial velocity {\it versus} frequency.
Radial velocities are measured by ultrasonic Doppler velocimetry along a radial beam shot from
a transducer at $-40 \deg$ in latitude. We then derive this image of color--coded log of power--spectral
density as a function of adimensional radius $r/a$ and frequency.
The rotation frequency of the sphere is $f = 4.3 \text{Hz}$ and the differential rotation frequency of
the inner sphere is $\Delta f = - 23 \text{Hz}$.
The power is high in bands at several frequencies, as in the spectrogram of figure \ref{fig:ddpwaves}.
}
        \label{fig:DOPwaves}
\end{figure}

Once again, useful information can be gained from ultrasonic Doppler velocity measurements. The bands that were first seen
on power spectra of the induced magnetic field are also present in the power spectra of velocity, as illustrated in figure
\ref{fig:DOPwaves}. They clearly extend deep into the sphere.

% ---------------- FIGURE DOP WAVES -----------------------------------------------------------
% -------------------------------------------------------------------------------------------
% ---------------- FIGURE DOP WAVES -----------------------------------------------------------
\begin{figure}
  \centerline{ \includegraphics[width=10cm]{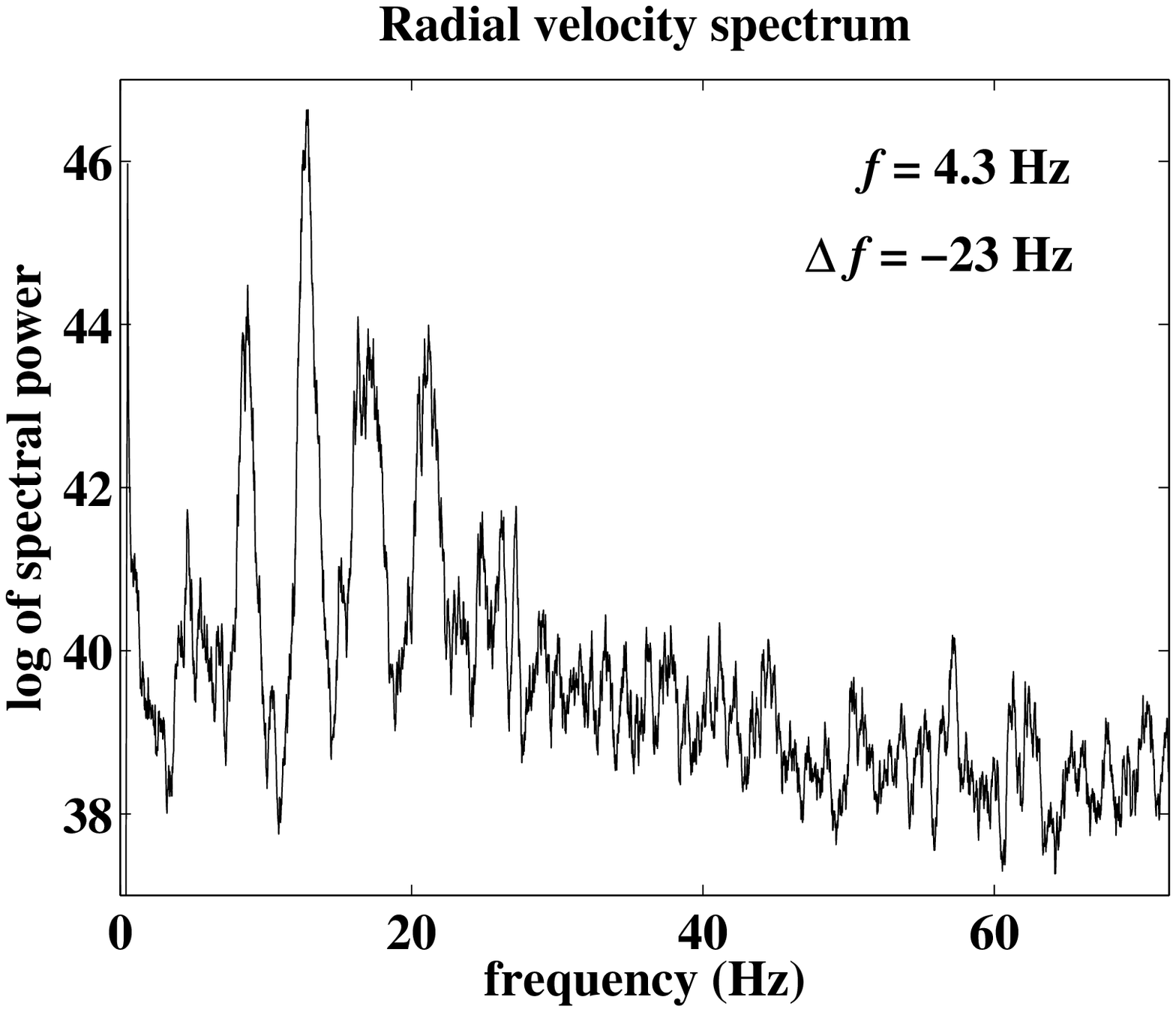}}
     \caption{Power spectral density of radial velocity {\it versus} frequency at mid--depth.
For the same record as in figure \ref{fig:DOPwaves}, we plot the log of power spectrum measured
at mid--depth. Peaks at several frequencies dominate the spectrum.
}
        \label{fig:DOP_spectrum}
\end{figure}
% -------------------------------------------------------------------------------------------

We have seen that, in the modified Taylor state, the maximum induced toroidal field is not near the inner sphere.
However, it is clear that a local perturbation of the azimuthal velocity near the inner sphere will induce a strong
azimuthal magnetic field, which will in turn change the magnetic field and the torque that drives the azimuthal flow.
It is tempting to see in this effect the origin of the waves we observe in $DTS$, but further work is needed to
corroborate this hypothesis.

%--------------------------------------------------------
\section{Implications for turbulence in the Earth's core}
\label{core_turbulence}
%--------------------------------------------------------

Our experiments illustrate that, due to the strong influence of the Coriolis force, motions are geostrophic when
the Elsasser number is less than 1, even though the magnetic torque on co--axial tubes is the only driving mechanism. 
Jault \cite{Jault08} argues that this property remains true even for large Elsasser numbers if one considers
time scales that are short compared to the Alfv\'en time. Pais and Jault \cite{Pais08} apply this idea
to retrieve geostrophic motions in the core that explain the observed secular variation of the magnetic
field of the Earth between years 2000 and 2005.

The idea that motions are organized in cylinders around the axis of rotation dates back to Taylor \cite{Taylor63}.
He pointed out that, in the steady--state, the magnetic torque acting on such cylinders must vanish
since, in the limit of small Ekman and Rossby numbers, there is no torque to balance it. The $DTS$ experiment
supports this idea, with the modification that the magnetic torque is balanced by friction at the core--mantle
boundary \cite{Kleeorin97}, and that this friction takes into account that the Ekman boundary layer becomes turbulent \cite{Nataf08}.
Although the local Reynolds number of the Ekman layers remain small in the core \cite{Desjardins04}, enhanced friction at the
core--mantle boundary can result from a rough boundary \cite{Lemouel06,Buffett07}, or from electromagnetic coupling
between the core and the mantle \cite{Buffett07}. In any case, nutation observations require a 10000--fold
enhancement of core--mantle coupling, as compared to linear Ekman friction \cite{Herring02}. It would be worth it to take
this effect into account in numerical simulations, in particular in quasi--geostrophic dynamo simulations \cite{Schaeffer06},
which are particularly efficient for exploring Earth--like fluid properties, and in which Ekman friction is already 
parameterized.

Our experiments also suggest that the only fluctuations left in the magnetostrophic regime, apart from localized turbulence
in the Ekman layers, are waves that propagate azimuthally in a retrograde fashion. These global wave phenomena are usually
excluded from the local turbulence analyses used to calibrate sub--grid algorithms in 3D numerical simulations \cite{Matsui07}.
This should probably be re--assessed once the origin of these waves is clearly established. In that respect, one should
note one important limitation of our $DTS$ experiment: the time--scales that characterize the various kinds of waves
are all very similar. For typical values of $f$ and $\Delta f$, the period of inertial waves and Alfv\'en waves, as well
as the dissipation time of Alfv\'en waves are all of the order of $0.1 \text{s}$. This makes it difficult to clearly identify the origin
of the waves, and it prevents a complete exploration of the various regimes.

It was soon recognized that departures from the Taylor state would trigger a special kind of Alfv\'en waves, called torsional oscillations \cite{Braginsky70}. Assuming reasonable values for the --hidden-- magnetic field within
the core, the period of these waves should be of the order of 10 years. Both core flow models derived from
the secular variation of the magnetic field \cite{Zatman97} and numerical simulations of the geodynamo \cite{Dumberry03} have been explored in order
to find evidence for these oscillations. The results indicate that part of
the fluid motions could indeed be torsional oscillations, providing a nice link with the observed length--of--day
variations \cite{Jault88}. However, recent analyses of the more precise satellite data also reveal large--scale motions
that are not torsional oscillations. Regional ``jerks'' are thus observed by Olsen and Mandea \cite{Olsen07}, while
the quasi--geostrophic core--flow models of Pais and Jault \cite{Pais08} are characterized by a large retrograde jet
that is completely offset with respect to the axis of rotation, in violation of the axisymmetric character of
torsional oscillations. All these observations, together with our detection of non--axisymmetric propagating waves
in the $DTS$ experiment suggest to us that the importance of torsional oscillations in the Earth's core could
have been overemphasized. However, we should note again that, because the Alfv\'en time in our experiment is of
the same order as the Joule dissipation time, torsional oscillations must be severely damped. It would be
of great value to conduct experiments where Joule dissipation is less important.

Finally, we emphasize that fluctuations are very minute when the flow is under the combined influence of a strong
magnetic field and rapid rotation (see figure \ref{fig:dopazi} and \ref{fig:ddpwaves}). Basically, the flow
is entirely constrained once the magnetic field and the rotation are given. When the Elsasser number is small, or when
the timescale of the motions is short \cite{Jault08}, the motions are constrained to be invariant along the rotation
axis (quasi--geostrophy) but are driven by the Lorentz forces. As magnetic diffusion is small at short time--scales,
only the flow can modify the magnetic field, and the same holds for the buoyancy field. One is thus led to conclude
that the turbulence at work in planetary cores has little to do with classical turbulence. Our $DTS$ experiment
is missing some of the key ingredients of a planetary dynamo, in fact it is {\it not} a dynamo. The $VKS$
experiment, which does behave as a dynamo \cite{Monchaux06}, is also missing key ingredients of planetary dynamos:
turbulence is almost entirely hydrodynamic and the produced magnetic field has a negligible effect on the flow. The
mechanisms of the spectacular
magnetic reversals observed in $VKS$ \cite{Berhanu07}, as those of the pioneer experiments of Lowes
and Wilkinson \cite{Lowes68}, probably have little to do with those at work in the Earth's core.

%-------------------
\section{Conclusion}
\label{conclusion}
%-------------------

Laboratory experiments such as our $DTS$ experiment help us better understand the peculiar organization of fluid flow and magnetic field at work in planetary dynamos. When the Coriolis and Lorentz forces are dominant, and the Elsasser number
less than 1, a modified Taylor state is observed, where the geometry is given by the rotation (invariance along the axis
of rotation) and the driving by the Lorentz force. In a Taylor state, the magnetic torque on co--axial tubes self--adjusts to
zero, since there is no balancing torque. In our experiments, as foreseen by Kleeorin {\it et al} \cite{Kleeorin97}, Ekman friction
at the external surface balances the magnetic torque. In this modified Taylor state, the flow remains geostrophic (azimuthal) and its
variation with cylindrical radius is entirely controlled by the balance between the magnetic torque the resulting shear produces and
the friction at the external boundaries. Using ultrasonic Doppler velocimetry, we measure time--averaged azimuthal velocity profiles in perfect agreement with this analysis.

Under the combined constraints of the imposed rotation and magnetic field, velocity fluctuations are very minute even though the
Reynolds number is in the range $10^6$. When present, the fluctuations reveal a very peculiar behavior: the power spectra are
dominated by peaks at several frequencies. The detailed analysis of the signals reveal that they correspond to waves that
propagate in a retrograde fashion with respect to the fluid \cite{Schmitt08}. The spectral peaks correspond to various azimuthal
wave numbers. It is not yet clear what are the driving forces of these magneto--inertial waves.

We infer that turbulence at work in planetary dynamos is of a very special kind: the flow is a complete slave of rotation and
magnetic forces, Ekman friction at the boundaries providing a balancing torque. But the magnetic field itself is advected by
the flow, and the same holds for the buoyancy field.

Two main limitations of the $DTS$ experiment prevent it of approaching regimes more relevant for planetary situations. The first one is that it is not a dynamo experiment, hence the magnetic field is not free to be advected by the flow. However, in most experimental
dynamos, the produced magnetic field is too weak to substantially alter the flow, in contrast with what occurs in $DTS$.
The second limitation is that all relevant time--scales (Alfv\'en, Joule dissipation, rotation, Rossby times) are in the same
range (of order $0.1 \text {s}$), in contrast with planetary situations. This makes it very difficult to isolate which are
the main forces controlling the propagating waves we observe, and extrapolating to natural situations. Building an experiment
exempt of these two limitations represents a tantalizing challenge, which has led Dan Lathrop and his group to build a promising 3m--diameter experiment in Maryland.

% \appendix

% \section{}
% \label{}

% The Acknowledgements are also a un-numbered section
%--------------------------
\section*{Acknowledgements}
% Acknowledgements text here
%--------------------------

We are thankful to all members of the geodynamo team in Grenoble for their contribution to the results and ideas presented in this article. We acknowledge useful comments from an anonymous reviewer. The $DTS$ project is supported by Fonds National de la Science, Institut National des Sciences de l'Univers, Centre National de la Recherche Scientifique, R\'egion Rh\^ one-Alpes and Universit\'e Joseph Fourier.

%--------------------------

\end{document}